\documentclass[journal,onecolumn]{IEEEtran}
\usepackage[table,xcdraw]{xcolor}
\usepackage{setspace} 
\usepackage{amsmath,amsfonts,amssymb,epsfig,epstopdf,url,array}

\usepackage{amsthm}
\usepackage{color,soul}
\usepackage{algorithmic}
\usepackage{balance}
\usepackage[ruled]{algorithm2e}
\usepackage{stmaryrd}
\usepackage{centernot}
\usepackage{tikz}
\usepackage{booktabs}
\usepackage{pdfpages}
\usepackage{graphicx}
\usepackage{subfigure}
\usepackage[font=small,skip=8pt]{caption}
\usepackage{tabularx}
\usepackage{comment}
\usepackage{multirow}
\usepackage{multicol}
\usepackage{bigstrut}
\usepackage{pdflscape}
\usepackage{rotating}
\usepackage{float}
\usepackage{lipsum}
\usepackage{enumitem}
\usetikzlibrary{plotmarks,shapes,arrows,chains,hobby,backgrounds,calc,trees}
\usepackage{makecell}
\usepackage{dblfloatfix}    
\usepackage{pdfcomment} 
\usepackage{arydshln}
\usepackage{amssymb}
\usepackage{bm}
\usepackage{arydshln}
\usepackage{pgfplots}
\usepackage{pgfplotstable}
\pgfplotsset{compat=newest}
\usepackage{lipsum}                     
\usepackage{xargs}                      
\usepackage[colorinlistoftodos,prependcaption,textsize=tiny]{todonotes}
\usepackage[utf8]{inputenc} 
\usepackage[T1]{fontenc} 
\usepackage{mathpazo} 
\usepackage[autostyle=true]{csquotes} 
\usepackage[section=subsection,sort=use]{glossaries}
\usepackage{datetime2,datetime2-calc}
\usepackage[utf8]{inputenc}
\usepackage[]{geometry}
\newcommandx{\unsure}[2][1=]{\todo[linecolor=red,backgroundcolor=red!25,bordercolor=red,#1]{#2}}
\newcommandx{\change}[2][1=]{\todo[linecolor=blue,backgroundcolor=blue!25,bordercolor=blue,#1]{#2}}
\newcommandx{\info}[2][1=]{\todo[linecolor=OliveGreen,backgroundcolor=OliveGreen!25,bordercolor=OliveGreen,#1]{#2}}
\newcommandx{\improvement}[2][1=]{\todo[linecolor=orange,backgroundcolor=orange!25,bordercolor=orange,#1]{#2}}
\newcommandx{\hiddencomment}[2][1=]{\todo[disable,#1]{#2}}

\definecolor{lightgray}{gray}{0.9}

\newtheoremstyle{exampstyle}
{0.5cm} 
{2\topsep} 
{} 
{} 
{\bfseries} 
{.} 
{.5em} 
{} 

\theoremstyle{exampstyle}

\theoremstyle{definition}

\theoremstyle{definition}
\newtheorem{mydef}{Definition}
\theoremstyle{definition}

\theoremstyle{definition}

\definecolor{header}{rgb}{0.0,0.0,0.0}
\definecolor{myblue}{rgb}{0.5,0.5,0.5}
\DTMnewdatestyle{Myyyy}{%
}
\DTMsetdatestyle{Myyyy}

\newglossarystyle{formel_altlong4colheader}{%
	\setglossarystyle{altlong4colheader}%
}
\newglossarystyle{abbreviationStyle}{%
	\setglossarystyle{long3col}%
	\renewenvironment{theglossary}%
	{\begin{supertabular}[l]{@{}p{0.4\hsize}p{0.9\hsize}p{0.01\hsize}@{}}}%
		{\end{supertabular}}%
}
\newglossarystyle{notationStyle}{%
	\setglossarystyle{long3col}%
	{\begin{supertabular}[l]{@{}p{0.4\hsize}p{0.9\hsize}p{0.01\hsize}@{}}}%
		{\end{supertabular}}%
}

\newglossary{abbreviation}{1i}{1o}{}
\newglossary{notation}{2i}{2o}{}
\newglossary{term}{3i}{3o}{}
\newglossaryentry{EV}{type=abbreviation,name=EV,description=Expected Value of a requirement subset}
\newglossaryentry{AV}{type=abbreviation,name=AV,description=Accumulated Value of a requirement subset}
\newglossaryentry{OV}{type=abbreviation,name=OV,description=Overall Value of a requirement subset}
\newglossaryentry{IP}{type=abbreviation,name=IP,description=Integer Programming}
\newglossaryentry{LP}{type=abbreviation,name=LP,description=Linear Programming}
\newglossaryentry{ILP}{type=abbreviation,name=ILP,description=Integer Linear Programming}
\newglossaryentry{MIP}{type=abbreviation,name=MIP,description=Mixed Integer Programming}
\newglossaryentry{DARS}{type=abbreviation,name=DARS,description=Dependency-aware Requirement Selection}
\newglossaryentry{DARS-IP}{type=abbreviation,name=DARS-IP,description=Integer Programming Method of DARS}
\newglossaryentry{DARS-ILP}{type=abbreviation,name=DARS-ILP,description=Integer Linear Programming Method of DARS}
\newglossaryentry{DARS-MIP}{type=abbreviation,name=DARS-MIP,description=Mixed Integer Programming Method of DARS}
\newglossaryentry{DARS-MIP-Combo}{type=abbreviation,name=DARS-MIP-Combo,description=The Combo Variation of DARS-MIP}
\newglossaryentry{DARS-MIP-Blind}{type=abbreviation,name=DARS-MIP-Blind,description=The Blind Variation of DARS-MIP}
\newglossaryentry{SDP}{type=abbreviation,name=SDP,description=Selection Deficiency Problem}
\newglossaryentry{BK}{type=abbreviation,name=BK,description=Binary Knapsack}
\newglossaryentry{BKP}{type=abbreviation,name=BKP,description=Binary Knapsack Problem}
\newglossaryentry{PCBK}{type=abbreviation,name=PCBK,description=Precedence Constrained Binary Knapsack}
\newglossaryentry{SBK}{type=abbreviation,name=SBK,description=Stochastic Binary Knapsack}
\newglossaryentry{NRP}{type=abbreviation,name=NRP,description=Next Release Problem}
\newglossaryentry{FRIG}{type=abbreviation,name=FRIG,description=Fuzzy Requirement Interdependency Graph}
\newglossaryentry{LOI}{type=abbreviation,name=LOI,description=Level Of Interdependency}
\newglossaryentry{RAN}{type=abbreviation,name=RAN,description=Radio Access Network}
\newglossaryentry{PMR}{type=abbreviation,name=PMR,description=Performance Management Recording}
\newglossaryentry{PMS}{type=abbreviation,name=PMS,description=Precious Messaging System}
\newglossaryentry{VDL}{type=abbreviation,name=VDL,description=Value Dependency Level}
\newglossaryentry{NVDL}{type=abbreviation,name=NVDL,description=Negative Value Dependency Level}
\newglossaryentry{PDL}{type=abbreviation,name=PDL,description=Precedence Dependency Level}
\newglossaryentry{NPDL}{type=abbreviation,name=NPDL,description=Negative Precedence Dependency Level}
\newglossaryentry{VDG}{type=abbreviation,name=VDG,description=Value Dependency Graph}
\newglossaryentry{PDG}{type=abbreviation,name=PDG,description=Precedence Dependency Graph}
\newglossaryentry{SPL}{type=abbreviation,name=SPL,description=Software Product Line}
\newglossaryentry{GRASP}{type=abbreviation,name=GRASP,description=Greedy Randomized Adaptive Search Procedures}
\newglossaryentry{TSP}{type=abbreviation,name=TSP,description=Target Sales Period}
\newglossaryentry{FIS}{type=abbreviation,name=FIS,description=Fuzzy Inference System}
\newglossaryentry{SRM}{type=abbreviation,name=SRM,description=Software Requirement Model}
\newglossaryentry{SRL}{type=abbreviation,name=SRL,description=Software Requirement List}
\newglossaryentry{PAPS}{type=abbreviation,name=PAPS,description=prioritization and partial selection}
\newglossaryentry{PAS}{type=abbreviation,name=PAS,description=Prioritization and Selection}
\newglossaryentry{GFG}{type=abbreviation,name=GFG,description=Goal-based Fuzzy Grammar}
\newglossaryentry{OBS}{type=abbreviation,name=OBS,description=Online Banking System}
\newglossaryentry{FCL}{type=abbreviation,name=FCL,description=Fuzzy Control Language}
\newglossaryentry{RDS}{type=abbreviation,name=RDS,description=Required Degree of Satisfaction}
\newglossaryentry{SVM}{type=abbreviation,name=SVM,description=Social Value Model}
\newglossaryentry{GQM}{type=abbreviation,name=GQM,description=Goal Question Metric}
\newglossaryentry{BKP-DIV}{type=abbreviation,name=BKP-DIV,description=Binary Knapsack Problem with Dependent Item Values}
\newglossaryentry{SKP}{type=abbreviation,name=SKP,description=Synergistic Knapsack Problem}
\newglossaryentry{OTKP}{type=abbreviation,name=OTKP,description=Oregon Trail Knapsack Problem}
\newglossaryentry{MV}{type=abbreviation,name=MV,description=Monetary Value}
\newglossaryentry{MVD}{type=abbreviation,name=SVD,description=Monetary Value Dependency}
\newglossaryentry{SVD}{type=abbreviation,name=SVD,description=Social Value Dependency}
\newglossaryentry{SORS}{type=abbreviation,name=SORS,description=Society-Oritented Requirement Selection}
\newglossaryentry{R}{type=notation,name=$R$,description=Set of requirements}
\newglossaryentry{r_i}{type=notation,name=$r_i$,description= Requirement $r_i \in R$}
\newglossaryentry{p(r_i)}{type=notation,name=$p(r_i)$,description= The probability that users select or use requirement $r_i$}
\newglossaryentry{v_i}{type=notation,name=$v_i$,description= Estimated value of a requirement $r_i$}
\newglossaryentry{vprime_i}{type=notation,name=$v^{\prime}_i$,description= Overall value of a requirement $r_i$}
\newglossaryentry{E(v_i)}{type=notation,name=$E(v_i)$,description= Expected value of a requirement $r_i$}
\newglossaryentry{c_i}{type=notation,name=$c_i$,description= Estimated cost of a requirement $r_i$}
\newglossaryentry{b}{type=notation,name=$b$,description= Available budget}
\newglossaryentry{x_i}{type=notation,name=$x_i$,description= Selection variable ($r_i \in R$ is selected or not)}
\newglossaryentry{O}{type=notation,name=$O$,description= Optimal subset}
\newglossaryentry{Otilde}{type=notation,name=$\tilde{O}$,description= Subset of excluded requirements}
\newglossaryentry{mu}{type=notation,name=$\tau$,description= Fuzzy membership function of requirements}
\newglossaryentry{rho}{type=notation,name=$\rho$,description= Fuzzy membership function of value dependencies}
\newglossaryentry{wedge}{type=notation,name=$\wedge$,description=Fuzzy AND operator}
\newglossaryentry{I_i}{type=notation,name=$I_{i}$,description= Impact of requirements on $r_i$}
\newglossaryentry{vee}{type=notation,name=$\vee$,description=Fuzzy OR operator}
\newglossaryentry{D}{type=notation,name=$D$,description=Set of explicit value dependencies}
\newglossaryentry{rho_infty_pos}{type=notation,name=$\rho^{+\infty}(r_i\text{,}r_j)$,description=Strength of all positive value dependencies from $r_i$ to $r_j$}
\newglossaryentry{rho_infty_neg}{type=notation,name=$\rho^{-\infty}(r_i\text{,}r_j)$,description=Strength of all negative value dependencies from $r_i$ to $r_j$}
\newglossaryentry{sigma}{type=notation,name=$\sigma(r_i\text{,}r_j)$,description=Specifies quality of a value dependency from $r_i$ to $r_j$}
\newglossaryentry{matrix_preference}{type=notation,name=$M_{n\times k}$,description= matrix of preferences for $n$ requirements and $k$ users}
\newglossaryentry{omega_low}{type=notation,name=$\omega_{-}$,description= The lower-bound of the confidence interval for Odds ratio}
\newglossaryentry{omega_high}{type=notation,name=$\omega_{+}$,description= The upper-bound of the confidence interval for Odds ratio}
\newglossaryentry{penalty}{type=notation,name=$\theta_{i,k}$,description= The penalty for the type $k$ value of a requirement $r_i$}
\newglossaryentry{priceLimit}{type=notation,name=$\gamma$,description= Price limit}
\newglossaryentry{P}{type=notation,name=$P$,description= The Set of Fuzzy Derivation Rules}
\newglossaryentry{G}{type=notation,name=$G$,description= The Set of Software Goals}
\newglossaryentry{g_j}{type=notation,name=$g_j$,description= Goal or subgoal $g_j \in G$}
\newglossaryentry{DC_g(x)}{type=notation,name=$DC_{g}(x)$,description= the Impact of the Requirement/Goal $ x $ on Satisfaction of the Goal $ g $}
\newglossaryentry{oplus}{type=notation,name=$\oplus$,description= Fuzzy OR operator (taking maximimum)}
\newglossaryentry{otimes}{type=notation,name=$\otimes$,description= Fuzzy AND operator (taking minimum)}
\newglossaryentry{zeta_i}{type=notation,name=$\zeta_i$,description= The estimated effort for complete satisfaction of $r_i$}
\newglossaryentry{zeta_prime_i}{type=notation,name=$\zeta_i^\prime$,description= The RELAX-ed effort for partial satisfaction of $r_i$}

\newglossaryentry{eta}{type=notation,name={$\eta$},description={Measure of causal strength}}

\makeglossaries

\begin{document}
\bibliographystyle{abbrv}

\title{An Integer Programming Model for\\ Embedding Social Values \\into Software Requirement Selection}

\author{Davoud~Mougouei\\ \small University of Wollongong, Australia\normalsize}


\maketitle

\onehalfspacing

\begin{abstract}
The existing software requirement selection methods have mainly focused on optimizing the economic value of a software product while ignoring its social values and their long-term impacts on the society. Social values however, are also important and need to be taken into account in software requirement selection. Moreover, social values of software requirements may change in the presence or absence of other requirements due to the \textit{value dependencies} among those requirements. These dependencies are imprecise and hard to specify in software projects. This paper presents an \textit{Integer Programming} (IP) model for the integration of social values and dependencies among them into software requirement selection. We further, account for the imprecision of social values and dependencies among them using the algebraic structure of fuzzy graphs. 
\end{abstract}

\vspace{-0.25cm}\hspace{0.75cm}\small\textbf{Keywords} Integer Programming, Social Values, Fuzzy, Software Requirement Selection \normalsize

\IEEEpeerreviewmaketitle

\section{Introduction}

\begin{figure*}[!htbp]
	\centering\includegraphics[scale=0.75]{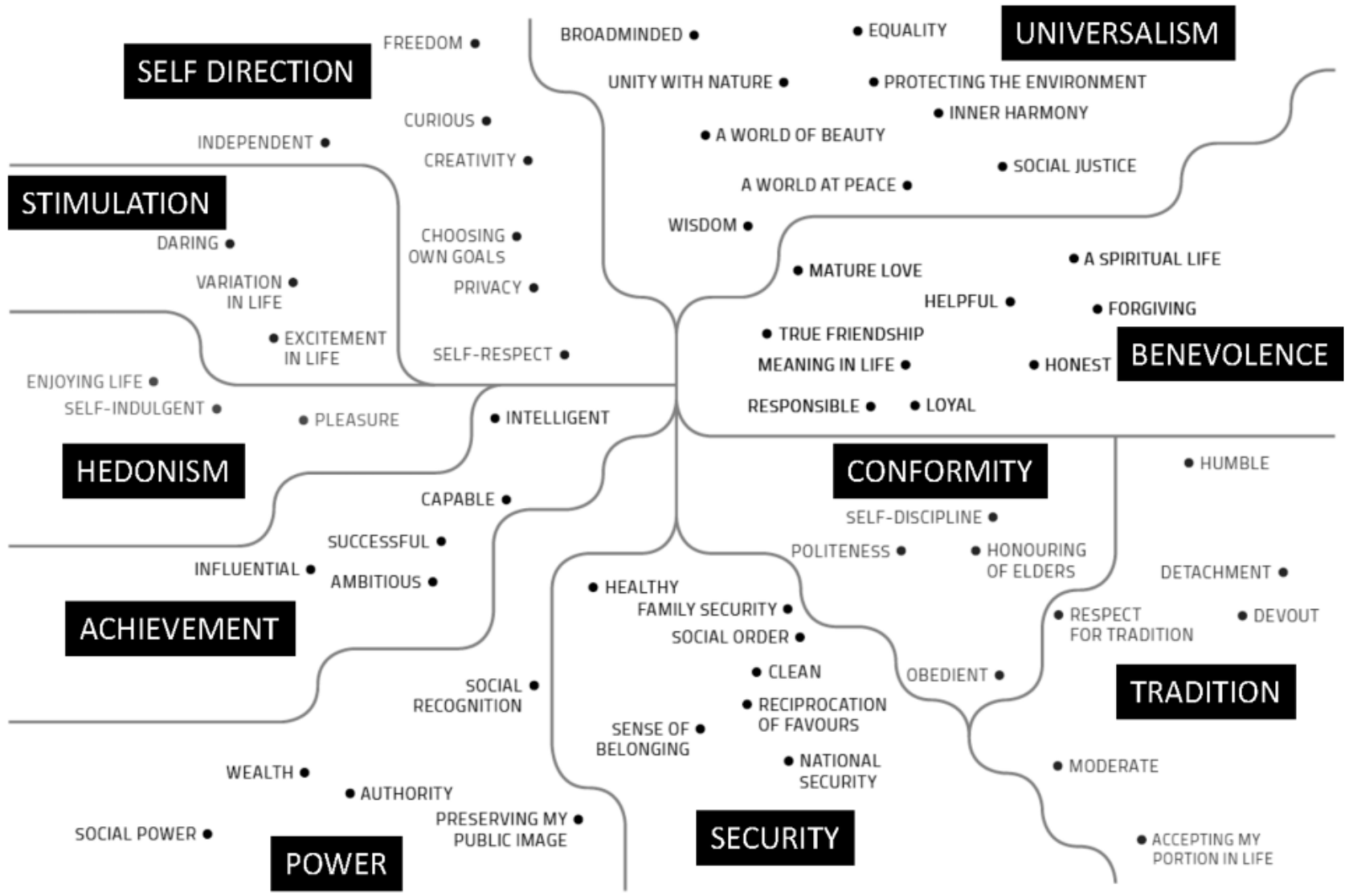}
	\vspace{0.5cm}
	\caption{%
		A map of social values appeared in~\cite{ferrario2016values}.
	}%
	\label{fig_map}
\end{figure*}

Software requirement selection~\cite{zhang2019software}, also known as \textit{Software Release Planning}~\cite{bagnall_next_2001,ameller2016survey,franch2016software}, is to find an optimal subset of requirements with the highest economic value while respecting the project constraints~\cite{dahlstedt2005requirements}. Thus, the optimization models of the existing requirement selection methods aim to optimize the economic value of a software product without caring for its social values and the long-term impacts of those values on the society~\cite{mougouei2018operationalizing,perera2019study,perera2019towards,hussain2018integrating}. To consider social values in software requirement selection, these values need to be integrated into the optimization models of requirement selection methods. A sample map of the economic and social values in software products is demonstrated in Figure~\ref{fig_map}.  

It is widely known that in a software project the economic values of the selected requirements may positively or negatively depend on the presence or absence of other requirements~\cite{mougouei2020dependency,mougouei2019dependency,mougouei2018mathematical,Zhang_RIM_2013,Robinson_RIM_2003,mougouei2016factoring,mougouei2017dependency} in the selected subset of requirements, i.e. \textit{Optimal Subset}. Analogously, there are also dependencies among social values of requirements in the sense that the presence or absence of certain requirements may impact the social values of other requirements due to the relationships and conflicts among those requirements. 

Hence, it is important that we take into account dependencies among social values as well as the dependencies among economic values of requirements in the optimization models of software requirement selection methods. Dependencies among social values and dependencies among economic values are all referred to as value dependencies for the ease of reference in this paper. Moreover, as observed by Carlshamre \textit{et al.}~\cite{carlshamre_industrial_2001}, requirement dependencies in general and value dependencies in particular are \emph{fuzzy}~\cite{carlshamre_industrial_2001} in the sense that the strengths those dependencies are imprecise and vary~\cite{dahlstedt2005requirements,ngo_wicked_2008,ngo2005measuring,carlshamre_industrial_2001} from large to insignificant~\cite{wang_simulation_2012} in real-world projects. Hence, it is important to consider not only the existence but the strengths of value dependencies and the imprecision of those dependencies in software projects. 

In this paper we present an integer programming method~\cite{mougouei2017integer} for considering the economic and social values with dependencies among those values in software requirement selection. In doing so, we have developed an optimization model that allows for embedding social values and dependencies among them into software requirement selection. The model is referred to as the \textit{Society-Oriented Requirement Selection} (\gls{SORS}). We have further made use of a variation of fuzzy graphs, referred to as the \textit{Value Dependency Graphs} (\gls{VDG}s), which was developed in our earlier work~\cite{mougouei2020dependency,mougouei2017modeling} for modeling value dependencies in the problem of binary knapsack with dependent item values. 
\section{Modeling Economic and Social Value Dependencies}
\label{modeling}

In this section we discuss modeling economic and social value dependencies by value dependency graphs (VDGs) introduced in~\cite{mougouei2017integer}. The algebraic structure of VDGs are used for computing the influences of the requirements of a software project on the social values of each other.   

\subsection{Value Dependency Graphs}
\label{ch_dars_modeling_vdg}
 
Value dependency graphs were initially presented in~\cite{mougouei2017modeling} for modeling value dependencies among items of a knapsack in the binary knapsack problem with dependent item values. In this paper we use VDGs for modeling economic and social value dependencies and their characteristics (quality and strength). A brief definition of VDG is provided in Definition~\ref{def_vdg}. 

\begin{mydef}
	\label{def_vdg}
	\textit{Value Dependency Graph} (VDG) is a signed directed fuzzy graph~\cite{Wasserman1994} $G=(R,\sigma,\rho)$ where, requirements $R:\{r_1,...,r_n\}$ constitutes the graph nodes. Also, qualitative function \gls{sigma} $\rightarrow \{+,-,\pm\}$ and the membership function $\rho: (r_i,r_j)\rightarrow [0,1]$ denote qualities and strengths of an explicit value dependency (edge of the graph) from $r_i$ to $r_j$ receptively. Moreover, $\rho(r_i,r_j)=0$ and $\sigma(r_i,r_j)=\pm$ specify the absence of any explicit value dependency from $r_i$ to $r_j$. 
\end{mydef}

\subsection{Economic and Social Value Dependencies in VDGs}
\label{ch_dars_modeling_vdg}

Definition~\ref{def_vdg_valuedepndencies} provides a more comprehensive definition of value dependencies that includes both explicit and implicit dependencies among requirements of a software product based on the algebraic structure of fuzzy graphs. 

\begin{mydef}
	\label{def_vdg_valuedepndencies}
	\textit{Value Dependencies}. 
	A value dependency in a value dependency graph $G=(R,\sigma,\rho)$ is defined as a sequence of requirements $d_i:\big(r(0),...,r(k)\big)$ such that $\forall r(j) \in d_i$, $1 \leq j \leq k$ we have $\rho\big(r(j-1),r(j)\big) \neq 0$. $j\geq 0$ is the sequence of the $j^{th}$ requirement (node) denoted as $r(j)$ on the dependency path. A consecutive pair $\big(r(j-1),r(j)\big)$ specifies an explicit value dependency. 
\end{mydef}

\begin{align}
\label{Eq_ch_dars_vdg_strength}
&\forall d_i:\big(r(0),...,r(k)\big): \rho(d_i) = \bigwedge_{j=1}^{k}\text{ }\rho\big(r(j-1),r(j)\big) \\
\label{Eq_ch_dars_vdg_quality}
&\forall d_i:\big(r(0),...,r(k)\big): \sigma(d_i) = \prod_{j=1}^{k}\text{ }\sigma\big(r(j-1),r(j)\big)
\end{align}

Equation (\ref{Eq_ch_dars_vdg_strength}) computes the strength of a value dependency $d_i:\big(r(0),...,r(k)\big)$ by finding the strength of the weakest of the $k$ explicit dependencies on $d_i$. Fuzzy operator $\wedge$ denotes Zadeh's~\cite{zadeh_fuzzysets_1965} AND operation (infimum). On the other hand, the quality (positive or negative) of a value dependency $d_i:\big(r(0),...,r(k)\big)$ is calculated by qualitative serial inference~\cite{de1984qualitative,wellman1990formulation,kusiak_1995_dependency} as given by (\ref{Eq_ch_dars_vdg_quality}) and Table~\ref{table_ch_dars_inference}. 
\vspace{0.25cm}
\begin{table}[!htb]
	\caption{Qualitative Serial Inference in VDGs.}
	\label{table_ch_dars_inference}
	\centering
\resizebox {0.3\textwidth }{!}{
	\begin{tabular}{cc|ccc}
		\toprule[1.5pt]
		\multicolumn{2}{r|}{\multirow{2}[1]{*}{ $\sigma\big(r(j-1),r(j),r(j+1)\big)$}} &
		\multicolumn{3}{c}{$\sigma\big(r(j),r(j+1)\big)$}
		\\
		\multicolumn{2}{r|}{} &
		$+$ &
		$-$ &
		$\pm$
		\bigstrut[b]\\
		\hline
		\multicolumn{1}{c}{\multirow{3}[1]{*}{$\sigma\big(r(j-1),r(j)\big)$}} &
		$+$ &
		$+$ &
		$-$ &
		$\pm$
		\bigstrut[t]\\
		\multicolumn{1}{c}{} &
		$-$ &
		$-$ &
		$+$ &
		$\pm$
		\\
		\multicolumn{1}{c}{} &
		$\pm$ &
		$\pm$ &
		$\pm$ &
		$\pm$
		\\
    \bottomrule[1.5pt]
	\end{tabular}%
	}

\end{table}

Let $D=\{d_1,d_2,..., d_m\}$ be the set of all value dependencies from $r_i \in R$ to $r_j \in R$ in a VDG $G=(R,\sigma,\rho)$, where positive and negative dependencies can simultaneously exist from $r_i$ to $r_j$. The strength of all positive value dependency from $r_i$ to $r_j$ is denoted by \gls{rho_infty_pos} and calculated by (\ref{Eq_ch_dars_ultimate_strength_positive}), that is to find the strength of the strongest positive dependency~\cite{rosenfeld_fuzzygraph_1975} from $r_i$ to $r_j$. Fuzzy operators $\wedge$ and $\vee$ denote Zadeh's~\cite{zadeh_fuzzysets_1965} fuzzy AND (taking minimum) and fuzzy OR (taking maximum) operations respectively. In a similar way, the strength of negative value dependency from $r_i$ to $r_j$ is denoted by \gls{rho_infty_neg} and calculated by (\ref{Eq_ch_dars_ultimate_strength_negative}).

\begin{align}
\label{Eq_ch_dars_ultimate_strength_positive}
&\rho^{+\infty}(r_i,r_j) = \bigvee_{d_m\in D, \sigma(d_m)=+} \text{ } \rho(d_m) \\[2pt]
\label{Eq_ch_dars_ultimate_strength_negative}
&\rho^{-\infty}(r_i,r_j) = \bigvee_{d_m\in D, \sigma(d_i)=-} \text{ } \rho(d_m) 
\end{align}

A brute-force approach to computing $\rho^{+\infty}(r_i,r_j)$ or $\rho^{-\infty}(r_i,r_j)$ needs to calculate the strengths of all paths from $r_i$ to $r_j$ which is of complexity of $O(n!)$ for $n$ requirements (VDG nodes). To avoid such complexity, we have formulated the problem of calculating $\rho^{+\infty}(r_i,r_j)$ and $\rho^{-\infty}(r_i,r_j)$ as a widest path problem (also known as the maximum capacity path problem~\cite{vassilevska_all_2007}) which can be solved in polynomial time by Floyd-Warshall algorithm~\cite{floyd_1962}. 

In this regard, we devised a modified version of Floyd-Warshall algorithm (Algorithm~\ref{alg_ch_dars_strength}) that computes $\rho^{+\infty}(r_i,r_j)$ and $\rho^{-\infty}(r_i,r_j)$ for all pairs of requirements $(r_i,r_j),\text{ }r_i,r_j \in R:\{r_1,...,r_n\}$ with the time bound of $O(n^3)$. For each pair of requirements $(r_i,r_j)$ in a VDG $G=(R,\sigma,\rho)$, lines $20$ to $37$ of Algorithm~\ref{alg_ch_dars_strength} find the strength of all positive value dependencies and the strength of all negative value dependencies from $r_i$ to $r_j$.

\begin{algorithm}
	\normalsize
	\caption{Calculating Strengths of Value Dependencies.}
	\label{alg_ch_dars_strength}
	\begin{algorithmic}[1]
		\REQUIRE VDG $G=(R,\sigma,\rho)$
		\ENSURE $\rho^{+\infty}, \rho^{-\infty}$
		\FOR{\textbf{each} $r_i \in R$}
		\FOR{\textbf{each} $r_j \in R$}
		\STATE $\rho^{+\infty}(r_i,r_j) \leftarrow \rho^{-\infty}(r_i,r_j) \leftarrow -\infty$ 
		\ENDFOR
		\ENDFOR
		\FOR{\textbf{each} $r_i \in R$}
		\STATE $\rho(r_i,r_i)^{+\infty} \leftarrow \rho(r_i,r_i)^{-\infty} \leftarrow 0$
		\ENDFOR
		\FOR{\textbf{each} $r_i \in R$}
		\FOR{\textbf{each} $r_j \in R$}
		\IF{$\sigma(r_i,r_j) = +$}
		\STATE $\rho^{+\infty}(r_i,r_j) \leftarrow \rho(r_i,r_j)$
		\ELSIF{$\sigma(r_i,r_j) = -$}
		\STATE $\rho^{-\infty}(r_i,r_j) \leftarrow \rho(r_i,r_j)$
		\ENDIF
		\ENDFOR
		\ENDFOR
		\FOR{\textbf{each} $r_k \in R$}
		\FOR{\textbf{each} $r_i \in R$}
		\FOR{\textbf{each} $r_j \in R$}
		\IF{$min\big(\rho(r_i,r_k)^{+\infty}, \rho(r_k,r_j)^{+\infty}\big) > \rho^{+\infty}(r_i,r_j)$}
		\STATE $\rho^{+\infty}(r_i,r_j) \leftarrow  min(\rho(r_i,r_k)^{+\infty}, \rho(r_k,r_j)^{+\infty})$
		\ENDIF
		\IF{$min\big(\rho(r_i,r_k)^{-\infty}, \rho(r_k,r_j)^{-\infty}\big) > \rho^{+\infty}(r_i,r_j)$}
		\STATE $\rho^{+\infty}(r_i,r_j) \leftarrow  min(\rho(r_i,r_k)^{-\infty}, \rho(r_k,r_j)^{-\infty})$
		\ENDIF
		\IF{$min\big(\rho(r_i,r_k)^{+\infty}, \rho(r_k,r_j)^{-\infty}\big) > \rho^{-\infty}(r_i,r_j)$}
		\STATE $\rho^{-\infty}(r_i,r_j) \leftarrow  min(\rho(r_i,r_k)^{+\infty}, \rho(r_k,r_j)^{-\infty})$
		\ENDIF
		\IF{$min\big(\rho(r_i,r_k)^{-\infty}, \rho(r_k,r_j)^{+\infty}\big) > \rho^{-\infty}(r_i,r_j)$}
		\STATE $\rho^{-\infty}(r_i,r_j) \leftarrow  min(\rho(r_i,r_k)^{-\infty}, \rho(r_k,r_j)^{+\infty})$
		\ENDIF
		\ENDFOR
		\ENDFOR
		\ENDFOR
	\end{algorithmic}
\end{algorithm}

\begin{align}
\label{Eq_ch_dars_influence}
I_{i,j} = \rho^{+\infty}(r_i,r_j)-\rho^{-\infty}(r_i,r_j) 
\end{align}

The overall strength of all positive and negative value dependencies from $r_i$ to $r_j$ is referred to as the \textit{Overall Influence} of $r_j$ on the value of $r_i$ and denoted by $I_{i,j}$. $I_{i,j}$ as given by (\ref{Eq_ch_dars_influence}) is calculated by subtracting the strength of all negative value dependencies from $r_i$ to $r_j$ ($\rho(r_i,r_j)^-\infty$) from the strength of all positive value dependencies from $r_i$ to $r_j$ ($\rho(r_i,r_j)^+\infty$). It is clear that $I_{i,j}\in[-1,1]$. $I_{i,j}>0$ states that $r_j$ influences the value of $r_i$ in a positive way whereas $I_{i,j}<0$ indicates that the ultimate influence of $r_j$ on $r_i$ is negative.  

\subsection{The Proposed Integer Programming Model}
\label{ch_dars_selection_ov}

We consider two types of values in our proposed optimization model (SORS). First is the economic value, which is manifested in terms of revenue/profit. Second is the class of social values, which includes all types of social values as depicted in Figure~\ref{fig_map}. For the sake of notational convenience we specify the economic value of a software requirement $r_i$ by $v_{i,1}$ while the social values of $r_i$ are specified by $v_{i,2}, ..., v_{i,k_u}$. $k_u$ gives the total number of values including the economic value. 

\begin{align}
\label{eq_ch_dars_penalty}
\nonumber
\theta_{i,k}= &\displaystyle \bigvee_{j=1}^{n} \bigg(\frac{x_j\big(\lvert I_{i,j,k} \rvert-I_{i,j,k}\big) + (1-x_j)\big(\lvert I_{i,j,k}\rvert+I_{i,j,k}\big)}{2}\bigg)&&=\\ 
&\displaystyle \bigvee_{j=1}^{n} \bigg(\frac{\lvert I_{i,j,k} \rvert + (1-2x_j)I_{i,j,k}}{2}\bigg),&& i=1,...,n \\
\label{eq_ch_dars_penalty_c1}
& x_j \in\{0,1\},\quad \quad \quad \quad \quad \quad \quad  \hspace{0.2em} && j=1,...,n 
\end{align}

In order to account for the impact of value dependencies on different types of value we use the algebraic structure of fuzzy graphs for computing the penalties of ignoring (selecting) positive (negative) value dependencies of a requirement on its economic/social values. For a type $k$ value (\ref{eq_ch_dars_penalty})-(\ref{eq_ch_dars_penalty_c1}) compute the penalty of ignoring (selecting) requirements with positive (negative) influence on the values of the selected requirements. \gls{penalty} in this equation denotes the penalty for the type $k$ value of a requirement $r_i$, $n$ denotes the number of requirements and $x_j$ specifies whether a requirement $r_j$ is selected ($x_j=1$) or not ($x_j=0$). Also, $I_{i,j,k}$, as in (\ref{Eq_ch_dars_influence}), gives the positive or negative influence of $r_i$ on the type $k$ value of $r_j$.  

We made use of the algebraic structure of fuzzy graphs for computing the influences of requirements on the values of each other as explained in Section~\ref{modeling}. Accordingly, $\theta_{i,k}$ is computed using the fuzzy OR operator which is to take supremum over the strengths of all ignored positive dependencies and selected negative dependencies of $r_i$ for its corresponding type $k$ value dependency graph. 

Equations (\ref{eq_sors})-(\ref{eq_sors_c11}) give our proposed integer programming model~\cite{nesterov2018lectures,antunes2016multiobjective}. In these equations, $x_i$ is a selection variable denoting whether a requirement $r_i$ is selected ($x_i=1$) or ignored ($x_i=0$). Also $\theta_{i,k}$ in (\ref{eq_ch_dars_penalty}) specifies the penalty for the type $k$ value of a requirement $r_i$, which is the extent to which the type $k$ value of $r_i$ is impacted by ignoring (selecting) requirements with positive (negative) influences on the value of $r_i$. Also, $k_u$ specifies the total number of value types including the economic value while $k_{1}$ specifies the economic value. 

Constraint (\ref{eq_sors_c1}) specifies all constraints related to social values, i.e. \textit{Social Constraints}, except for the economic value, which is embedded into the objective function. $\alpha_k$ in (\ref{eq_sors_c1}) denotes the required lower bound for each social value. Finding proper values for $\alpha_k$ is specially important for reconciling conflicts among social values when satisfaction of one value conflicts with satisfaction of another one. $\alpha_k$ can be modified in such cases to suit value preferences of stakeholders.  

Constraint (\ref{eq_sors_c2}) ensures that the total cost of the requirements does not exceed the project budget $b$. Also, (\ref{eq_sors_c3}) in the proposed optimization model accounts for precedence dependencies among requirements and the value implications of those dependencies, which may impact all value types of values. Precedence dependencies mainly include requirement dependencies of type \textit{Requires}, where one requirement intrinsically requires the other one, and \textit{Conflicts-With}, where one requirement intrinsically conflicts with the other one.   

\begin{align}
\label{eq_sors}
&\text{Maximize }  \sum_{i=1}^{n} x_i v_{i,1} - y_i v_{i,1}\\[5pt]
\label{eq_sors_c1}
&\text{Subject to} \sum_{i=1}^{n} x_i v_{i,k} - y_i v_{i,k} \geq \alpha_{k},& k=2,...,k_u\\[5pt]
\label{eq_sors_c2}
&\sum_{i=1}^{n} c_i x_i \leq b\\[5pt]
\label{eq_sors_c3}
& \begin{cases}
x_i \le x_j  & r_j \text{ precedes } r_i \\[5pt]
x_i \le 1-x_j& r_i \text{ conflicts with } r_j,\text{ }i\neq j= 1,...,n
\end{cases}\\[5pt]
\label{eq_sors_c4}
& \theta_{i,k}\geq \bigg(\frac{\lvert I_{i,j,k} \rvert + (1-2x_j)I_{i,j,k}}{2}\bigg),& i\neq j = 1,...,n,\text{ }k=1,...,k_{u} \\
\label{eq_sors_c5}
& -g_i \leq x_i \leq  g_i,& i=1,...,n\\[5pt]
\label{eq_sors_c6}
& 1-(1-g_i) \leq x_i \leq 1+(1-g_i),& i=1,...,n\\[5pt]
\label{eq_sors_c7}
& -g_i \leq y_i \leq g_i,& i=1,...,n\\
\label{eq_sors_c8}
& -(1-g_i)\leq(y_i-\theta_i) \leq (1-g_i),& i=1,...,n\\[5pt]
\label{eq_sors_c9}
&\text{ } 0 \leq y_i \leq 1,& i = 1,...,n\\[5pt]
\label{eq_sors_c10}
&\text{ } 0 \leq \theta_i \leq 1,& i = 1,...,n \\[5pt]
\label{eq_sors_c11}
& \text{ }x_i,g_i \in \{0,1\},& i=1,...,n
\end{align}

In (\ref{eq_sors})-(\ref{eq_sors_c11}) we have either $a:(x_i=0,y_i=0)$, or $b:(x_i=1,y_i=\theta_{i,k})$ occur. To capture the relation between $\theta_{i,k}$ and $y_i$ in a linear form, we have made use of an auxiliary variable $g_i=\{0,1\}$ and (\ref{eq_sors_c5})-(\ref{eq_sors_c11}). As such, we have either $(g_i=0) \rightarrow a$, or $(g_i=1) \rightarrow b$. The selection model (\ref{eq_sors})-(\ref{eq_sors_c11}) therefore is a linear model as it has a linear objective function with linear inequality constraints constraints. 

\section{Summary and Future Work}

This paper presented an integer programming model, referred to as SORS (Society-Oriented Requirement Selection), for taking into account social values as well as the dependencies among those values in software requirement selection. The model is linear and therefore scalable to software projects with large number of requirements. We used the algebraic structure of fuzzy graphs for capturing the imprecision of value dependencies. 

This paper can be extended in several directions. Identification of dependencies among social values, for instance, is one important research direction as value dependencies serve as the input for our proposed requirement selection model. Measuring social values is also another important aspect of this research, which is particularly required for specifying the lower-bounds of the social constraints in the optimization model of the SORS. Moreover, it is important to develop such identification and measurement techniques in full participation with stakeholders of software products as social values may change across different classes of stakeholders. In this regard, establishing a collaborative platform for participatory development of standards for identification and measuring social values can be of significant benefit.

\small
\bibliography{ref}

\end{document}